\documentclass[conference, onecolumn]{IEEEtran}
\IEEEoverridecommandlockouts
\usepackage{svg}
\usepackage{amsmath, amsfonts,amssymb,amsthm, bm} 
\usepackage{mathtools}
\usepackage{graphicx}
\usepackage{float}
\usepackage{mathtools}
\usepackage{xcolor}
\usepackage{harpoon}
\usepackage{array}
\usepackage{tabularx}
\usepackage{algorithm,algcompatible}
\usepackage{algpseudocode}
\usepackage{algorithm}
\allowdisplaybreaks[1]
        \usepackage{setspace}

\algnewcommand\INPUT{\item[\textbf{Input:}]}%
\algnewcommand\OUTPUT{\item[\textbf{Output:}]}%
\newcommand{\bb}{\mathbf}

\usepackage{caption}
\usepackage{subcaption}

\def\BibTeX{{\rm B\kern-.05em{\sc i\kern-.025em b}\kern-.08em
    T\kern-.1667em\lower.7ex\hbox{E}\kern-.125emX}}
    
\begin{document}

\title{Robust Singular Values based on L1-norm PCA\\
}

\author{\IEEEauthorblockN{Duc H. Le$^\ddag$ and Panos P. Markopoulos$^{\dag *}$\thanks{$^*$Corresponding author.}}
\IEEEauthorblockA{$^\ddag$\textit{Dept. of Electr. \& Microelectron. Engineering}, 
\textit{Rochester Institute of Technology}, 
Rochester, NY \\
E-mail: dhl3772@rit.edu}
\IEEEauthorblockA{$^\dag$\textit{Depts. of Electr. \& Comput. Engineering and Comput. Science}, 
\textit{The University of Texas at San Antonio}, 
San Antonio, TX \\
E-mail: panos@utsa.edu}
}

\maketitle

\begin{abstract}
Singular-Value Decomposition (SVD) is a ubiquitous data analysis method in engineering, science, and statistics. Singular-value estimation, in particular, is of critical importance in an array of engineering applications, such as channel estimation in communication systems, electromyography signal analysis, and image compression, to name just a few. Conventional SVD of a data matrix coincides with standard Principal-Component Analysis (PCA). The L2-norm (sum of squared values) formulation of PCA promotes peripheral data points and, thus, makes PCA sensitive against outliers. Naturally, SVD inherits this outlier sensitivity. In this work, we present a novel robust non-parametric method for SVD and singular-value estimation based on a L1-norm (sum of absolute values) formulation, which we name L1-cSVD. Accordingly, the proposed method demonstrates sturdy resistance against outliers and can facilitate more reliable data analysis and processing in a wide range of engineering applications.
\end{abstract}

\section{Introduction}

Singular-Value Decomposition (SVD) has established itself as a powerful tool, ubiquitous in various engineering applications. For example, applying SVD to the channel matrix of a multiple-input multiple-output (MIMO) channel decomposes the MIMO channel into multiple single-input single-output (SISO) channels with gains corresponding to singular values, which enables efficient power allocation and channel capacity estimation \cite{lebrun:2005, fisher:2002}. Furthermore, SVD has been extensively employed in various watermarking schemes \cite{chung:2007,chang:2005}, direction-of-arrival (DOA) estimation \cite{malioutov:2005,markopoulos:2014:doa}, restructuring of deep neural network acoustic models \cite{xue:2013}, electromyography (EMG) signal analysis \cite{iqbal:2017}, etc.

Another data analysis method that is closely related to SVD is Principal-Component Analysis (PCA), which is also useful in a number of fields, such as machine learning, signal processing, and pattern recognition \cite{jolliffe:1986,bishop:2006,duda:2001}. The traditional PCA method seeks to maximize the L2-norm of the variance of the projected coordinates on the principal components (PCs). However, because of its emphasis on the square of the coordinates, PCA is sensitive against corruption from gross and sparse outliers. 

Thus, there has been considerable research effort in reformulating PCA employing the L1-norm instead (L1-PCA), which is able to suppress the effect of extreme data points \cite{brooks:2013,ke:2005}. The exact solution to L1-PCA can be obtained in polynomial time with respect to the number of data points \cite{markopoulos:2014}. However, optimality can be traded for lower computational complexity, which has been implemented in a greedy approach \cite{kwak:2008}, a semidefinite programming approach \cite{mccoy:2011}, an alternating algorithm \cite{nie:2011}, and a bit-flipping algorithm \cite{markopoulos:2017}, just to name a few. Apart from L1-PCA, another line of research that aims to ameliorate the effect of outliers corruption is Robust Principal-Component Analysis (RPCA) \cite{candes:2011}, which strives to decompose a matrix into a sparse and a low-rank component. 

On the other hand, besides the PCs, or equivalently the left singular vectors of SVD \cite{meyer:2000}, a robust, outlier-resistant acquisition of singular values (SVs) is also of great interest. Regrettably, L1-PCs, while robust against outliers, do not possess the attractive property of their L2-norm counterpart to diagonalize the data matrix $\bb{X}$ \cite{gang:2022}, thus making the extension from L1-PCA to SVs estimation non-trivial.

In this paper,  we leverage the robustness of the subspace found by previous L1-PCA algorithms to apply to SVs estimation. To this end, we propose an algorithm named L1-cSVD that finds SVs and right singular vectors from a given L1-subspace by solving a re-orthogonalization problem. We then test the performance of this L1-cSVD algorithm with synthetic and real dataset against the state-of-the-art RPCA, which corroborates the robustness of the proposed algorithm in preserving SVs when facing outliers. 

%-----------------------
\section{Technical Background} \label{tech background}
\subsection{Standard SVD and PCA} \label{SVD and PCA}
SVD decomposes a $D\times N$ matrix $\bb{X}$ as \cite{meyer:2000}
\begin{align}
    {\bf X = U \Sigma V}^T,
\end{align}
where ${\bb{U}}\in \mathbb{R}^{D\times d}$ and ${\bb{V}} \in \mathbb{R}^{N\times d}$ are orthonormal matrices, defined as the left and right singular vectors respectively, ${\bb \Sigma} \in \mathbb{R}^{d\times d}$ is a positive-valued diagonal matrix whose diagonal elements are the singular values (SVs), and $d = {\rm rank}(\bb{X})$. This is the ``compact" SVD (cSVD) where the left and right singular vectors corresponding to zero SVs are disregarded \cite{meyer:2000}. For simplicity, we will refer to ``compact" SVD as SVD throughout this paper. 

Standard SVD is very closely related to PCA, since the first $K\,(K\leq d)$ left singular vectors of $\bf{U}$ are also the first $K$ PCs of $\bb{X}$, maximizing the L2-norm projection \cite{jolliffe:1986}
\begin{align} \label{proj max L2}
    \bb{Q}_{L2} = \underset{\bb{Q} \in \mathbb{S}^{D\times K} }{\rm argmax} {||\bb{Q}^T \bb{X}||}_{2,2},
\end{align}
where $\mathbb{S}^{D\times K}$ denotes the set of orthonormal matrices in $\mathbb{R}^{D\times K}$ (Stiefel manifold) and $||\cdot||_{2,2}$ denotes the Frobenius or L2-norm of its matrix argument \cite{meyer:2000}.
%-------------------------------------------------------
\subsection{L1-PCA} \label{past l1pca}
The proposed method builds on top of L1-PCA, which is mathematically formulated by replacing the L2-norm in the optimization problem of (\ref{proj max L2}) with the L1-norm (sum of absolute values), as
\begin{align} \label{proj max L1}
    {\bf Q}_{L1} = \underset{\bb{Q} \in \mathbb{S}^{D\times K}}{\rm argmax} ||\bb{Q}^T \bb{X}||_{1,1}.
\end{align}
L1-PCA can be extended to robust SVs estimation by taking the standard SVD of the projected matrix $\bb{Q}_{L1}\bb{Q}_{L1}^T\bb{X}$. The optimal solution to (\ref{proj max L1}) was presented for the first time in \cite{markopoulos:2014} and has polynomial cost in $N$. In this work, we focus on suboptimal approaches with lower time complexity. 

\subsubsection{Greedy Algorithm with Successive Nullspace Projection}
Kwak \cite{kwak:2008} proposed an iterative algorithm to solve (\ref{proj max L1}) when $K=1$. The algorithm can be summarized as
\begin{align}
    \bb{b}^{(t)} = {\rm sgn} \left(\bb{X}^T \bb{X} \bb{b}^{(t-1)} \right),
\end{align}
$t = 2,3,4,...$, where $\bb{b}^{(1)} \subset \{\pm 1\}^N$ is an antipodal binary vector that can be randomly initialized. Then, the PC can be approximated to be $\bb{q} = \bb{Xb}/||\bb{Xb}||_{2}$. For $K>1$, the PCs of $\bb{Q}$ are found in a greedy way, by replacing $\bb{X}$ with its projection onto the nullspace of the previously found PCs. %The time complexity of this algorithm is $\mathcal{O} (MDNK)$ where $M$ is the maximum number of iterations. If $M$ is considered to be bounded by $N$, the complexity becomes $\mathcal{O}(DN^2K)$  \cite{kwak:2008,markopoulos:2017}. 
It is important to note that because L1-PCA is not scalable, meaning the PCs themselves are dependent on the number of PCs being found, the greedy approach is suboptimal. %less optimal than finding the PCs $\bb{q}_i$ jointly.

\subsubsection{Iterative Alternating Algorithm}
Nie et al. \cite{nie:2011} presented a method that finds the column vectors of $\bb{Q}$ jointly. The iterative algorithm can be summarized as
\begin{align}
\bb{B}^{(t)} = {\rm sgn} (\bb{X}^T \bb{Q}^{(t-1)}), \bb{Q}^{(t)} = \mathcal{U}(\bb{XB}^{(t)}),
\end{align}
$t = 2,3,4,...$, where $\mathcal{U}(\cdot)$ returns the closest orthonormal matrix using the Procrustes theorem \cite{meyer:2000} and $\bb{B}^{(1)} \subset \{\pm 1\}^{N\times K}$ is a binary matrix that can be arbitrarily initialized. %The complexity of this algorithm is $\mathcal{O}[M(DN+K^2)]$ where $M$ is the maximum number of iterations. Consider $M$ to be of the same order of magnitude as NK, the complexity becomes $\mathcal{O} (DN^2K + NK^3)$ \cite{nie:2011,markopoulos:2017}.

\subsubsection{Bit-flipping Algorithm}
Markopoulos et al. \cite{markopoulos:2017} proposed an algorithm that calculates the effect of flipping any single bit of the binary matrix $\bb{B}$ on the optimization metric of (\ref{proj max L1}) and flips the bit that yields the highest increase to the metric. The algorithm converges to the optimal L1-PCs with high frequency and frequently achieves higher value in the optimization metric of (\ref{proj max L1}) than previous alternatives. %The complexity of this algorithm is $\mathcal{O}(ND\min\{N,D\}+N^2K^2(K^2+d))$ where $d = {\rm rank}(\bb{X})$.

\subsection{RPCA}
Another line of research to the problem of robustly recovering a low-rank structure from a corrupted matrix is RPCA \cite{candes:2011}. This approach seeks to decompose a matrix $\bb{X}$ into a low-rank component $\bb{L}$ and a sparse component $\bb{S}$ that models sparse outliers by solving the problem
\begin{align} \label{rpca formulation}
    \underset{\bb{L, S, L+S = X}}{\rm minimize} ||\bb{L}||_* + \lambda ||\bb{S}||_{1,1},
\end{align}
where $(\cdot)_*$ indicates the nuclear norm (sum of singular values). The problem essentially promotes the sparsity of $\bb{S}$ by minimizing its L1-norm and the sparsity of the SVs of $\bb{L}$ by minimizing the nuclear norm of $\bb{L}$ or equivalently the L1-norm of its SVs, thus making $\bb{L}$ low-rank. The performance of RPCA depends largely on $\lambda$ \cite{chen:2009}. In this work, we set $\lambda = 1/\sqrt{M}$ where $M$ is the larger dimension of $\bb{X}$ \cite{candes:2011}. RPCA can be extended to SVs estimation by taking the conventional SVD of the extracted low-rank component, i.e., $(\bb{U}, \bb{\Sigma}, \bb{V}) = {\rm SVD} (\bb{L})$.
%--------------------------------------
\section{Proposed Method} \label{proposed solution}
%Naively, as soon as the L1-PCs are found by solving Eq.~(\ref{proj max L1}) using one of the mentioned method, one may perform conventional SVD on the low-rank approximated data matrix $\bb{Q}_{L1} \bb{Q}_{L1}^T \bb{X} = \bb{U}_{L1} \bb{\Sigma}_{L1} \bb{V}^T_{L1}$. However, the SVs $\bb{\Sigma}_{L1}$ found using this method is very close to the SVs found by the conventional L2-cSVD and thus are not robust against outliers, which calls for a more sophisticated methods to find new $(\bb{\Sigma}_{L1}, \bb{V}_{L1})$ combinations.

\subsection{Proposed Algorithm: L1-cSVD}
We now formulate our L1-norm based SVD approach to be
\begin{align} \label{L1SVD formulation}
    \bb{X} \approx \bb{U}_{L1} \bb{\Sigma}_{L1} \bb{V}^T_{L1},
\end{align}
where the left and right singular vectors $\bb{U}_{L1}$ and $\bb{V}_{L1}$, respectively, are orthonormal and $\bb{\Sigma}_{L1}$ is diagonal. As a result, this decomposition has to be an approximated one because the only exact decomposition with such constraints on $\bb{U}_{L1}, \bb{\Sigma}_{L1}$ and $\bb{V}_{L1}$ would be the conventional SVD due to its uniqueness property. 

We call our algorithm L1-``compact" SVD (L1-cSVD) to emphasize that we only collect $K\leq D\leq N$ SVs and singular vectors from $\bb{X}$. We carry over the property of SVD that the left singular vectors are also the PCs and set $\mathbf{U}_{L1}$ to the L1-PCs $\mathbf{Q}_{L1}$ obtained by solving (\ref{proj max L1}).  This choice of $\bb{U}_{L1}$ ensures that the subspace found is robust against outliers \cite{markopoulos:2014} \cite{markopoulos:2017}. As mentioned in section \ref{past l1pca}, the problem in Eq.~(\ref{proj max L1}) has been studied extensively and there are multiple algorithms available to choose from, the importance of which will be discussed in detail in section \ref{importance U}. For now, we assume that a good $\bb{U}_{L1}$ can be found. In this section, we propose an algorithm to find the SVs $\bb{\Sigma}_{L1}$ from the left singular vectors $\bb{U}_{L1}$.

Conventional SVD has an attractive property which states that the left singular vectors $\bb{U}_{L2} = \bb{Q}_{L2}$ from (\ref{proj max L2}) also diagonalize $\bb{X}$; i.e., $\bb {U}_{L2}^T \bb{X} = \bb{\Sigma}_{L2} \bb{V}_{L2}^T$ is an orthogonal matrix or equivalently $\bb{U}_{L2}^T \bb{XX}^T \bb{U}_{L2}$ is diagonal, and it is the only orthonormal matrix with this property \footnote{In this paper, ``orthogonal matrix" means a matrix whose column vectors are orthogonal but not necessarily normalized}. Thus, an orthonormal $\bb{U}_{L1}$ generally cannot diagonalize $\bb{X}$, while the formulation of L1-cSVD in (\ref{L1SVD formulation}) requires that $\bb{\Sigma}_{L1} \bb{V}_{L1}^T$ is orthogonal. As a result, the problem of L1-cSVD becomes finding the closest orthogonal matrix $\bb{\Sigma}_{L1} \bb{V}_{L1}^T$ to $\bb{U}_{L1}^T \bb{X}$ using the L1-norm,
\begin{align} \label{S, V L1}
    (\bb{\Sigma}_{L1}, \bb{V}_{L1}) = \underset{ \substack{{\bb{V}} \in \mathbb{S}^{N\times K,}\\\bb{\Sigma} \in {\rm diag}(\mathbb{R}^{K})}}
    {\rm argmin} {||\bb{X}^T\bb{U}_{L1} - \bb{V}\bb{\Sigma}||}_{1,1}.
\end{align}
This is a non-convex problem due to the orthonormality constraint on $\bb{V}$ \cite{boyd}. We will solve for the matrices $\bb{\Sigma}$ and $\bb{V}$ suboptimally by an alternating method. For fixed $\bb{V}$, finding $\bb{\Sigma}$ can be equivalently decomposed into $K$ individual problems

\begin{align}
    \underset{\sigma_i \in \mathbb{R}}{\rm minimize} {||(\bb{X}^T\bb{U}_{L1})_{:,i} - \sigma_{i} \bb{v}_{i}||}_{1},
\end{align}
$i = 1,2,...,K$, where $\sigma_i$ is the $i^{\rm th}$ SV in $\bb{\Sigma}$. This problem is simply seeking a scaling factor $\sigma_{i}$ that minimizes the L1-distance between vectors $(\bb{X}^T\bb{U}_{L1})_{:,i}$ and $\sigma_{i} \bb{v}_{i}$. The answer is found in \cite{markopoulos:2016} to be $\sigma_i = (\bb{X}^T\bb{U}_{L1})_{j_{\rm opt},i}/v_{j_{\rm opt}, i}$ where
\begin{align} \label{find Simga L1}
    j_{\rm opt} = \underset{j\in \{1,2,...,N\}}{\rm argmin} {\Bigg|\Bigg|(\bb{X}^T\bb{U}_{L1})_{:,i} - \frac{(\bb{X}^T\bb{U}_{L1})_{j,i} }{v_{j,i}} \bb{v}_{i}\Bigg|\Bigg|}_{1}&,
\end{align}
which performs exhaustive search on $N$ candidates for $\sigma_i$ chosen such that $(\bb{X}^T\bb{U}_{L1})_{:,i}$ is equal to $\sigma_{i} \bb{v}_i$ on the $j^{\rm th}$ entry. From the $N$ candidates, the one that returns the least L1 error will be chosen to be $\sigma_i$. On the other hand, for fixed $\bb{\Sigma}$, $\bb{V}$ is found by solving
\begin{align} 
    \underset{\bb{V}\in \mathbb{S}^{N\times K}}{\rm minimize} {||\bb{X}^T\bb{U}_{L1} - \bb{V}\bb{\Sigma}||}_{1,1},
\end{align}
which is essentially an L1-norm Orthogonal Procrustes problem. A solution to this problem using a smoothed version of the L1-norm has been studied in \cite{trendafilov:2003}. However, in this paper, for lower computational complexity, we will use the solution to the L2-Orthogonal Procrustes problem instead, since it is empirically observed that given the L1-informed $\bb{U}_{L1}$ and $\bb{\Sigma}_{L1}$, the L1 solution for $\bb{V}$ gives similar result to its L2 counterpart while taking much longer to solve. Thus, we set $\bb{V}= \bb{U'V'}^T$ where ${\bb{(U', \Sigma', V')} }= {\rm SVD}(\bb{X}^T \bb{U}_{L1} \bb{\Sigma}^{-1})$. This result for $\bb{V}$ is then used to update the SVs $\bb{\Sigma}$, which is used to refine $\bb{V}$ in an alternating fashion until convergence. Upon termination of the alternating updates, $\bb{V}_{L1} = \bb{V}$ and $\bb{\Sigma}_{L1} = \bb{\Sigma}$. The algorithm can be summarized in the pseudocode in Algorithm~1.

\begin{algorithm} [t!]
    \caption{L1-cSVD (proposed)} \label{L1 algo}
  \begin{algorithmic}[1] 
    \INPUT Data matrix $\bb{X}_{D\times N}$, number of SVs $K$
    \State $\bb{U} \leftarrow$ L1PCA $(\bb{X})$
    \State $\bb{A} \leftarrow \bb{X}^T \bb{U}$
    \State \textbf{initialization} $\bb{\Sigma} \leftarrow$  zeros(K,K), orthonormal $\bb{V}$ 
    \While{not converged}
    \For{i = 1 to K}
        \For{j = 1 to N}
            \State $\rm s_j \leftarrow \rm([\mathbf{A}]_{j,i}/[\mathbf{V}]_{j,i})$
            \State$\rm M_{j} \leftarrow ||[\bb{A}]_{:,i} - s[\bb{V}]_{:,i}||_1 $
        \EndFor
        \State $\rm j_{\rm opt} \leftarrow \underset{j\in [1:N]}{\rm argmin} \{M_j\}$ 
        \State$\rm[\bb{\Sigma}]_{i,i} \leftarrow s_{j_{\rm opt}}$
    \EndFor
    \State $(\bb{U}', \bb{\Sigma}', \bb{V}') \leftarrow$  SVD($\bb{A\Sigma}^{-1}$)
    \State $\bb{V} \leftarrow \bb{U}'\bb{V}'^T$
    \EndWhile
  \OUTPUT {$\bb{U}_{L1} \leftarrow \bb{U}, \bb{\Sigma}_{L1}\leftarrow \bb{\Sigma}, \bb{V}_{L1} \leftarrow \bb{V} $}
    \end{algorithmic}
\end{algorithm}

At this point, it is worth noting that finding $||\bb{A}_{:,i} - s\bb{V}_{:,i}||_1$ costs $\mathcal{O}(N)$ for a candidate $s$. Since there are $N$ candidates for $K$ SVs, finding $\bb{\Sigma}_{L1}$ costs $\mathcal{O}(N^2K)$ in total. $\bb{V}$ is found with cost $\mathcal{O}(DK^2)$. Because $N\geq D\geq K$, the complexity of finding $\bb{\Sigma}_{L1}$ and $\bb{V}_{L1}$ is $\mathcal{O}(WKN^2)$, where $W$ is the number of iterations. By considering $W$ to be bounded by $NK$, the complexity of this L1-cSVD algorithm is $\mathcal{O}(N^3K^2)$ in addition to the cost of the L1-PCA algorithm chosen to find $\bb{U}_{L1}$.
\subsection{Importance of Choosing Left Singular Vectors $\bb{U}_{L1}$: Joint vs Greedy} \label{importance U}
As previously mentioned, the L1-PCA problem of Eq.~(\ref{proj max L1}) finding the left singular vectors $\bb{Q}_{L1}=\bb{U}_{L1}$, on which $\bb{\Sigma}_{L1}$ and $\bb{V}_{L1}$ are dependent, can be solved by Greedy solutions \cite{kwak:2008} or Joint solutions \cite{nie:2011,markopoulos:2017}. Because L1-PCA is not scalable, the Joint solutions have a more optimal $||\bb{Q}_{L1}^T \bb{X}||_{1,1}$ metric. 

However, since maximizing the L1-norm promotes balance among the columns of $\bb{Q}_{L1}^T \bb{X}$, Joint L1-PCA also inadvertently rotates the L1-PCs along the L1-subspace to make the projections more balanced. Therefore, their L1-PCs are less aligned to the L2-PCs than those found by Greedy L1-PCA, which finds individual L1-PCs without having to balance with other L1-PCs. %On the other hand, this issue is ameliorated by the Greedy solution since it focuses on maximizing the L1-norm of the projection to one particular PC .

Coincidentally, since the paper is concerned with robust SVs estimation, finding good PCs should be given a priority to finding a good subspace because the SVs are directly tied to their corresponding PCs. As a result, we elect to choose the Greedy solution in \cite{kwak:2008} to find $\bb{Q}_{L1}$ in Eq.~(\ref{proj max L1}).
%====================================================================================
\section{Experimental Studies} \label{experimental studies}
\subsection{Algorithm Analysis: Convergence}
\begin{figure} 
\centering\includegraphics[width=0.5\linewidth]{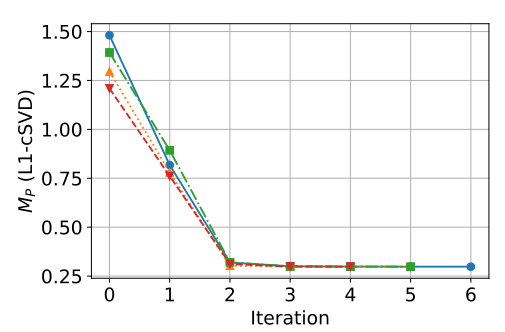}  
\caption{Evolution of the performance metric $M_P$ for the L1-cSVD algorithm for 4 distinct random initializations of $\bb{V}$.} \label{convg L1}
\end{figure}

To assess the convergence of the L1-cSVD algorithm, we define the normalized performance measurement $M_P = ||\bb{U}^T \bb{X} - \bb{\Sigma V}^T||_{1,1}/||\bb{U}^T \bb{X}||_{1,1}$ and plot its evolution for 4 different initializations on the same $8\times50$ data matrix $\bb{X}$ ($K = 5$ SVs are obtained) in Fig.~\ref{convg L1}. We see that for all 4 initializations, L1-cSVD converges to the same value in just 6 iterations. 
%------------------------------------------------------------------------
\subsection{Performance Analysis with Synthetic Dataset} \label{performance analysis}
In this section, we will compare the SVs estimation criterion of the L1-cSVD algorithm against the conventional SVD and RPCA.
\subsubsection{Signal Model}
We consider a clean data matrix $\bb{X}^{\rm clean} \in \mathbb{R}^{D\times N}$ of rank-K $(K \leq D \leq  N)$, containing data from the subspace spanned by $\bb{U}_0\in \mathbb{S}^{D\times K} $, which is kept constant for the experiment. $\bb{V}_0 \in \mathbb{S}^{N\times K}$ is a random orthonormal matrix and the SVs $\bb{\Sigma}_0$ are drawn from a log-uniform distribution. First, $\bb{X}^{\rm clean}$ is corrupted by Gaussian noise $\bb{N}$ with a signal-to-noise ratio ${\rm SNR} = ||\bb{N}||_{2,2}^2/||\bb{\Sigma}_0||_{2,2}^2$. Then, the noisy data matrix is further corrupted by matrix $\bb{O}$ containing outliers from a subspace spanned by $\bb{R}_o\in \mathbb{S}^{D\times K_o} $, which is also kept constant for the experiment. The probability of corruption is $P_o$, so $\bb{\Gamma} \in \{0,1\}^{D\times N}$ has $P_o$ chance of a column vector being $\bb{1}$ while the rest are $\bb{0}$. The entries of $\bb{S}_o \in \mathbb{R}^{K_o \times N}$ are drawn from a zero-mean normal distribution with variance chosen to attain a certain outlier-to-signal ratio, defined as ${\rm OSR} = ||\bb{O}||_{2,2}^2/||\bb{\Sigma}_0||_{2,2}^2$.
\begin{align}
    \bb{X^{\rm corrupted}} &= \bb{X}^{\rm clean} + \bb{N} + \bb{O} \nonumber \\
    &= \bb{U}_0 \bb{\Sigma}_0 \bb{V}_0^T + \bb{N} + \bb{\Gamma} \odot \bb{R}_o \bb{S}_o.
\end{align}
%-------------------------------------------------------------
\subsubsection{Singular-Value Preservation}
We then define the normalized SVs estimation error metric to evaluate how well different algorithms preserve SVs under subspace outlier corruption
\begin{align}
    R_{\rm sv} = \frac{||\bb{\Sigma}^{\rm estimated} - \bb{\Sigma}^{\rm clean}||_{2,2}}{|| \bb{\Sigma}^{\rm clean}||_{2,2}},
\end{align}
where $\bb{\Sigma}^{\rm clean}$ is calculated by applying the conventional SVD on the clean data matrix $\bb{X}^{\rm clean}$ and $\bb{\Sigma}^{\rm estimated}$ is the estimated SVs from the corrupted dataset $\bb{X}^{\rm corrupted}$ by applying different SVD algorithms.
\begin{figure}
\centering\includegraphics[width=0.6\linewidth]{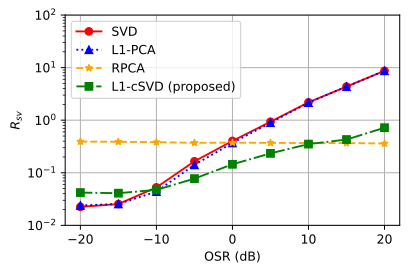}  
\caption{The normalized total SVs error $R_{\rm sv}$ for different SVs estimation approaches at different OSR dB values averaged over 1000 experiments. The synthetic dataset has $D = 10$, $N =50, K = 4$ SVs are captured, and $\rm SNR = 10$ dB. Outliers have $P_o = 0.04$ and $K_o = 4$.} \label{figure SV}
\end{figure}

In Fig.~\ref{figure SV}, we observe that L1-cSVD is just as good as SVD when outliers are not present. However, when the data is corrupted, L1-cSVD attains superior performance with a normalized SVs estimation error of $5-25$\% for OSR between -10dB and 5dB, in which the SVs estimated by SVD starts to deviate strongly. In addition, L1-cSVD follows SVD very closely at low OSR, indicating that it is in effect the same as SVD at this regime. 

L1-PCA has the same SVs estimation error as standard SVD, indicating that the extension from robustness of PCs to robustness of SVs is not trivial. RPCA has the same performance at every OSR due to its ability to separate the sparse component effectively. However, the reconstructed low-rank component is not necessarily robust, since its SVs estimation error is about 40\%. This can be attributed to $\bb{X}^{\rm corrupted}$ being noisy because the formulation in (\ref{rpca formulation}) does not take into account noise, which is neither low-rank nor sparse. 

We are also interested in comparing the preservation of the individual SVs, defined by
\begin{align}
    &R_{\rm sv,i} = \frac{(\sigma_i^{\rm estimated} - \sigma_i^{\rm clean})^2}{ (\sigma_i^{\rm clean})^2},
\end{align}
$i = 1,2,...,K$.
\begin{figure} 
\centering\includegraphics[width=0.8\linewidth]{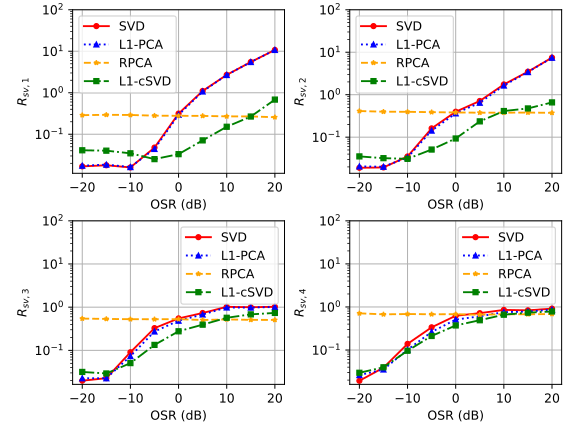} 
\caption{The normalized first, second, third and fourth SV errors $R_{\rm sv,1}$, $R_{\rm sv,2}$, $R_{\rm sv,3}$, and $R_{\rm sv, 4}$, respectively, for different SVs estimation methods.} \label{fig SVi}
\end{figure}

The most significant finding is that the robustness of L1-cSVD is clearly demonstrated in the estimation of the first and also most important SV, which only deviates less than 10\% from the clean dominant SV for OSR up to almost 10dB. On the other hand, RPCA incurs a constant 30-70\% error on all SVs across different OSR values. 

In addition, the performance of RPCA is strongly dependent on the parameter $\lambda$, which might make it fail when data is corrupted by just benign noise. On the other hand, the proposed method L1-cSVD is non-parametric and thus does not suffer from the same issue. 

%The estimation of subdominant SVs by L1-cSVD is less robust because the task of finding subdominant SVs is more challenging than the first SV, explainable by the fact they they are tied to the subdominant L1-PCs. According to the method of Greedy L1-PCA \cite{kwak:2008}, if the first L1-PC is not recovered perfectly, the second L2-PC will likely to not be recovered perfectly either because it has to be orthogonal to the 1st L1-PC, which is not readily orthogonal to the 2nd clean PC as the 1st and 2nd clean PCs are already orthogonal. Nevertheless, L1-cSVD still consistently attains lower SV-estimation error than SV. 
%=============================================================
\subsection{Bayesian Classifier}
In this section, we apply the robust L1-cSVD method to the Bayesian classifier problem. We choose the ``Vowel" dataset from Penn Machine Learning Benchmarks (PMLB) \cite{olson:2017}. The chosen dataset has $C = 11$ vowels to be classified, $N = 990$ samples evenly divided among the 11 classes and $D = 11$ numerical features available. From 90 samples of each of the 11 vowels, we use 75 for training and reserve 15 for testing. 

We apply a Bayesian Classifier on each training dataset $\bb{X}^{(i)}$ of the $i^{\rm th}$ vowel, $i = 1,2,...,11$, where we find the median vector $\bb{m}_i \in \mathbb{R}^{D}$ (taken instead of the conventional mean vector for better outlier resistance), and the SVD of $\bb{X}^{(i)} =\bb{U}^{(i)} \bb{\Sigma}^{(i)}\bb{V}^{(i)T} $. %Here, $\bb{U}^{(i)}$ and $\bb{\Sigma}^{(i)}$ are also the eigenvectors and the square root of the eigenvalues of the covariance matrix of $\bb{X}^{(i)}$. 
Thus, given a test data point $\bb{y} \in \mathbb{R}^{D}$, its Mahalanobis distance to the distribution of the $i^{\rm th}$ vowel is \cite{duda:2001}
\begin{align}
    d_i = \sqrt{\sum_{j = 1}^{D} \Bigg(\frac{\bb{u}^{(i)T}_j (\bb{y} - \bb{m}^{(i)})}{\sigma^{(i)}_j/\sqrt{N}}\Bigg)^2}.
\end{align}
According to the Bayesian Classifier \cite{duda:2001}, $\bb{y}$ is classified to the class with the smallest $d_i$.

To assess the robustness of L1-cSVD, we then additively corrupt 3 out of the 75 entries of the training dataset $\bb{X}^{(i)}$ with outliers of 25 times the average power of an entry of $\bb{X}$. As a reference to our synthetic data experiment, the OSR in this case can be calculated to be 0 dB; i.e., the outlier has the same energy as the clean data.

\begin{figure} 
\centering\includegraphics[width=0.85\linewidth]{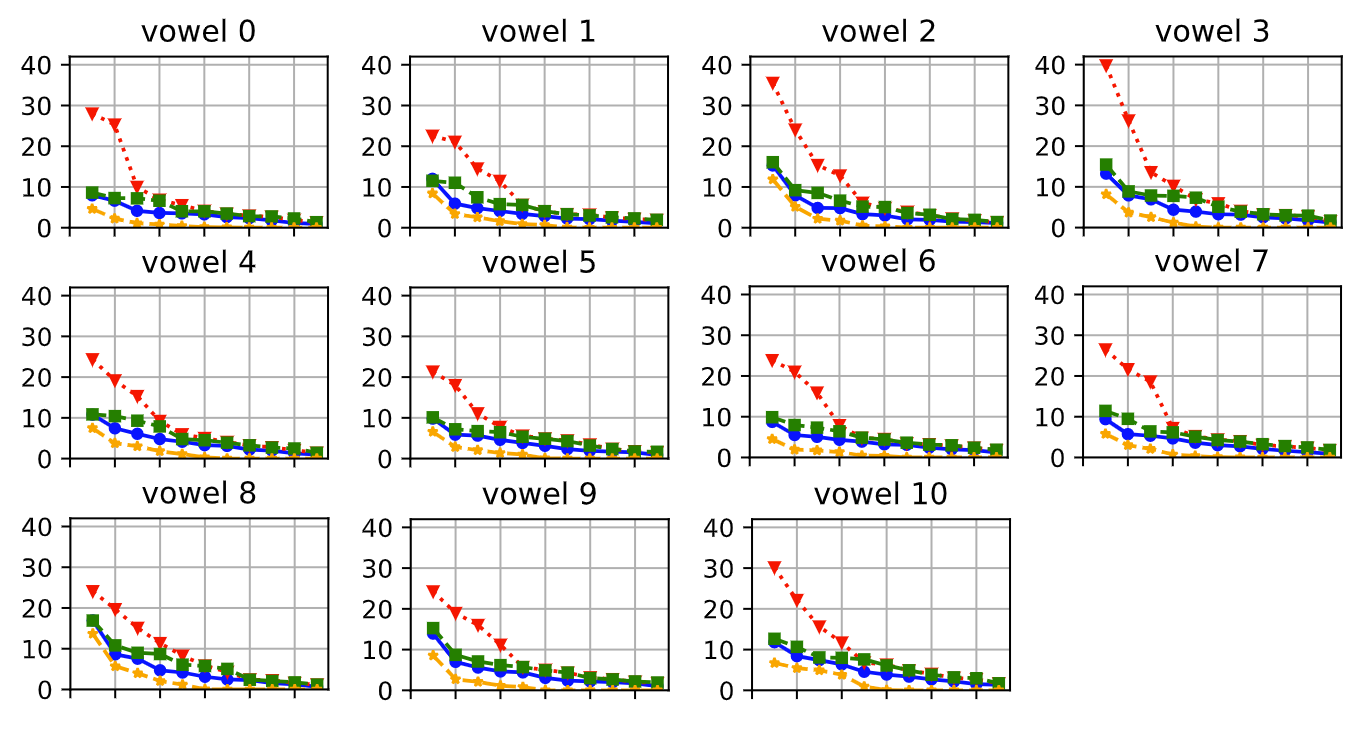} 
\caption{The ground truth SVs of each vowel training dataset for clean data using SVD (blue, circles) along with the estimated SVs for corrupted training data using SVD (red, triangles), RPCA (orange, stars) and L1-cSVD (green, squares).} \label{SV vowels}
\end{figure}

We then classify the vowels based on the parameters trained by the corrupted training data using $\bb{U}^{(i)}$ and $\bb{\Sigma}^{(i)}$ from either SVD, RPCA, or L1-cSVD and compare the trained parameters. In Fig.~(\ref{SV vowels}), it can be clearly seen that L1-cSVD is able to reconstruct the SVs very well from corrupted data compared to the traditional SVD. Importantly, the first or dominant SVs are reconstructed almost exactly by L1-cSVD. On the other hand, RPCA tends to underestimate the SVs, possibly because the sparsity of SVs is overly promoted. 

\begin{figure} [H]
\centering\includegraphics[width=0.55\linewidth]{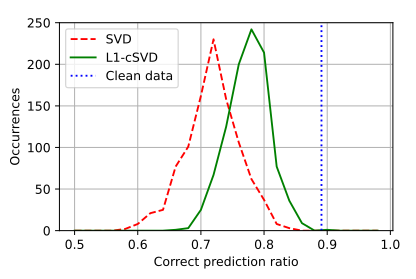} 
\caption{Histogram of for the correct prediction ratio when the corrupted data is trained with SVD (green) and L1-cSVD (red, dashed) for 1000 experiments with different corruption realizations. The correct prediction ratio using SVD on clean data is marked by the blue dotted line as a benchmark.} \label{histogram vowels}
\end{figure}

Thus, we proceed to train the corrupted dataset with L1-cSVD and compare its performance to applying SVD on corrupted and clean data. In Fig.~(\ref{histogram vowels}), we can observe that L1-cSVD attains a higher correct prediction ratio than the conventional SVD, demonstrating its robustness against gross and sparse outliers. 

%----------------------------------------

\subsection{Direction-of-Arrival Estimation}

In this experiment, we choose a linear array with $M = 8$ sensors uniformly spaced by $\lambda/2$ taking $T = 200$ snapshots of 3 incoming signals with directions of arrival (DOAs) $-45^\circ, 0^\circ$ and $60^\circ$. The received signal can be written to be $\bb{Y} = \bb{AS} + \bb{N}$ \cite{malioutov:2005}, where $\bb{Y} \in \mathbb{C}^{M\times T}$ describes the signal of interest (SoI) received at a sensor at a time snapshot, $\bb{A} \in \mathbb{C}^{M\times N_\theta}$ is the array manifold matrix with $a_{m,k} = \exp \left[-j m \pi \sin(\theta_k) \right]$, and $\bb{N}$ is noise with SNR = 10 dB. The DOAs grid is chosen to be from $-90^\circ$ to $90^\circ$ with $1^\circ$ spacing, so $N_\theta = 180$. Because L1-cSVD is developed for real data, $\bb{A}$ and $\bb{Y}$ are realified to be $\Tilde{\bb{A}}\in \mathbb{R}^{2M\times N_{\theta}}$ and $\Tilde{\bb{Y}} \in \mathbb{R}^{2M\times T}$ by concatenating their real and imaginary components \cite{markopoulos:2014:doa,markopoulos:2014}.

The task of DOA estimation is to reconstruct $\bb{S} \in \mathbb{R}^{N_\theta\times T}$, which describes the amplitude of the incoming signals from $N_\theta$ DOAs at $T$ time snapshots. To better distinguish between spatially close sources, Malioutov et. al \cite{malioutov:2005} proposed a method that enforces sparsity within every column of $\bb{S}$, since signal sources can be considered sparse in space but not in time, using the L12-norm. In addition, to reduce time complexity, a dimensionality reduction preprocessing step is used, where $\Tilde{\bb{Y}}$ is replaced with $\Tilde{\bb{Y}}^{\rm SV}=\bb{U}_K \bb{\Sigma}_K$, where $\bb{U}_K$ are the first $K$ left singular vectors and $\bb{\Sigma}_K$ are the first $K$ SVs of $\Tilde{\bb{Y}}$, with $K = 3$ being the expected number of sources. Thus, we solve
\begin{align} 
    \underset{\bb{S}^{\rm SV} \in \mathbb{C}^{N_\theta \times K}}{\rm minimize} || \bb{\Tilde{Y}}^{\rm SV} - \Tilde{\bb{A}}\bb{S}^{\rm SV} ||_{2,2} + \lambda||\bb{S}^{\rm SV}||_{1,2},
\end{align}
where $\lambda$ is a regularization parameter, to obtain the robust spatial spectrum. This method is called $\ell 1$-SVD \cite{malioutov:2005}, since it uses an L1-norm to enforce spatial sparsity and SVD for dimensionality reduction. It is not to be confused with L1-cSVD in this work, which formulates and solves a compact SVD scheme using the L1-norm. 
Unfortunately, using conventional SVD for dimensionality reduction means that $\bb{Y}^{\rm SV}$ can be sensitive to outliers. To demonstrate this, we corrupt $\bb{Y}$ with jammer signals coming from DOAs $-30^\circ, 30^\circ$ and $50^\circ$. Each jammer corrupts 10 time snapshots at random with power of 20 times the SoI power (OSR = 0 dB). We reconstruct $\bb{S}^{\rm SV}$ from the corrupted received signal using the $\ell 1$-SVD method in \cite{malioutov:2005} with either SVD and L1-cSVD used for dimensionality reduction for comparison.

\begin{figure} [H]
    \centering \includegraphics[width=0.36\linewidth]{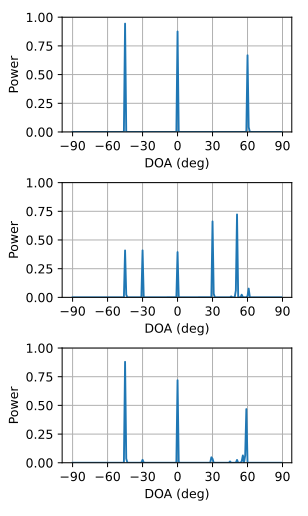}
    \caption{DoA spectra produced by the $\ell 1$-SVD method \cite{malioutov:2005} with (top) no jammers, (middle) jammers on, using conventional SVD for dimensionality reduction, and (bottom) the same jammers, using L1-cSVD for dimensionality reduction.}  \label{doa fig}
\end{figure}

In Fig.~\ref{doa fig}, it can be observed that by using L1-cSVD for dimensionality reduction before applying $\ell 1$-SVD for DOAs estimation, the jammers' peaks in the spectrum are effectively suppressed compared to using conventional SVD. %The reconstructed power of the 3 signals of interest (height of the peaks) is also well-preserved by using L1-cSVD for preprocessing.

%===========================================================================
\section{Conclusions} \label{conclusion}
We presented a novel algorithm for signal value decomposition, based on L1-PCA. This algorithm is the first one to extend the robustness against outliers of L1-PCA problem to finding robust SVs. We showed that such an extension is not obvious and proposed a problem formulation to find more robust SVs. We solved this problem by means of the proposed non-parametric L1-cSVD algorithm, which utilizes the L1-PCA basis $\bb{U}_{L1}$ and reorthogonalizes $\bb{U}_{L1}^T\bb{X}$ to find $\bb{\Sigma}_{L1}$ and $\bb{V}_{L1}$ with additional complexity $\mathcal{O}(N^3K^2)$. Our algorithm was tested on SVs estimation, Bayesian Classification, and DoA estimation, on both real and synthetic data. All experiments demonstrate that the proposed L1-cSVD algorithm is more robust in SVs estimation against corruption from sparse and gross outliers compared to the conventional SVD and RPCA, while maintaining similar performance to SVD for clean data. %In addition, being non-parametric, L1-cSVD can be generalized to a wider class of problems than RPCA. 
%\appendices
%\section{}

%\section*{Acknowledgment}


% \ifCLASSOPTIONcaptionsoff
%   \newpage
% \fi

\bibliographystyle{ieeetr}
\bibliography{arx_main}

\end{document}